\documentclass[secnumarabic,amssymb, nobibnotes, preprint, aps, longbibliography]{revtex4-1}
\usepackage{color}

\usepackage{amsmath}
\usepackage{graphicx}

\begin{document}
\title{Equivalence of magnetic field and particle dilution in the strange metal state of CeCoIn$_5$}
\author{Nikola Maksimovic}
\altaffiliation{Contact for correspondence, nikola\_maksimovic@berkeley.edu or analytis@berkeley.edu}
\affiliation{Department of Physics, University of California, Berkeley, California 94720, USA}
\affiliation{Materials Science Division, Lawrence Berkeley National Laboratory, Berkeley, California 94720, USA}

\author{Ian M. Hayes}
\affiliation{Department of Physics, University of California, Berkeley, California 94720, USA}
\affiliation{Materials Science Division, Lawrence Berkeley National Laboratory, Berkeley, California 94720, USA}

\author{Sooyoung Jang}
\affiliation{Department of Physics, University of California, Berkeley, California 94720, USA}
\affiliation{Materials Science Division, Lawrence Berkeley National Laboratory, Berkeley, California 94720, USA}

\author{Bayan Alizadeh}
\affiliation{Department of Physics, University of California, Berkeley, California 94720, USA}
\affiliation{Materials Science Division, Lawrence Berkeley National Laboratory, Berkeley, California 94720, USA}

\author{Ehud Altman}
\affiliation{Department of Physics, University of California, Berkeley, California 94720, USA}

\author{James G. Analytis$^*$}
\affiliation{Department of Physics, University of California, Berkeley, California 94720, USA}
\affiliation{Materials Science Division, Lawrence Berkeley National Laboratory, Berkeley, California 94720, USA}

\begin{abstract}
The Bardeen-Cooper-Schrieffer mechanism for superconductivity is a triumph of the theory of many-body systems. Implicit in its formulation is the existence of long-lived (quasi)particles, originating from the electronic building blocks of the materials, which interact to form Cooper pairs that move coherently in lock-step. The challenge of unconventional superconductors is that it is not only unclear what the nature of the interactions are, but whether the familiar quasi-particles that form a superconducting condensate even exist. In this work, we reveal, by the study of applied magnetic field in electronically diluted materials, that the metallic properties of the unconventional superconductor CeCoIn$_5$ are determined by the degree of quantum entanglement that (Kondo) hybridizes local and itinerant electrons. This work suggests that the properties of the strange metallic state are a reflection of the disentanglement of the many-body state into the underlying electronic building blocks of the system itself.

\end{abstract}

\maketitle

The normal state of a large class of correlated materials, including high-$T_c$ superconductors, is often called a ``strange'' metal. This understates the situation somewhat, as the most notable features, a resistivity that varies linearly with temperature as $T \rightarrow 0$ and a strongly temperature dependent Hall effect, both represent significant deviations from the conventional Fermi-liquid picture of metals~\cite{Landau1956}. An origin for these anomalous properties is still to be identified, but it is thought that the key is in the proximity of these materials to a transition, known as a Mott transition, that localizes electrons. Such transitions are associated with strong correlations and often entangle degrees of freedom that are normally independent in the conventional picture, and this can be revealed by the coupled response to distinct kinds of perturbations, for example, doping and pressure~\cite{Klintberg2010,Ferreira2008,Seo2015}. Recently, this type of coupled response was extended to strange metal transport in an iron-pnictide high-$T_{c}$ compound as a function of magnetic field and temperature~\cite{Hayes2016}. In this study, we reveal a simple connection between the effect of an applied field and that of doping on the magneto-transport in CeCoIn$_5$, suggesting that the origin of the anomalous transport properties of the strange metal arise directly from the degree of entanglement between local and itinerant electrons.

Heavy fermion materials such as $\text{CeCoIn}_{5}$ are composed of a lattice of 4$f$-electrons from the Ce ion, embedded in a sea of conduction ($c-$)electrons. The dominant interactions in the system have opposing tendencies; Mott-like localization of the $f$-electrons favoring magnetic order on the one hand~\cite{Koitzsch2008}, and on the other hand Kondo-driven hybridization of the $c/f$ electrons, favoring $f$-electron delocalization and the formation of a heavy Kondo hybridized band~\cite{Doniach1977}. A number of experiments have suggested this system is nearly localized, and that the hybridized band is not fully formed~\cite{Nakajima2007,Chen2017,VanDyke2014,Meyer2000,Pepin2007,Steglich2016}. This proximity to Mott-like localization of 4$f$-electrons may be the origin for both the strange metal behavior and the superconducting ground state~\cite{Chen2017,Nakatsuji2004,VanDyke2014,Steglich2016}, but precisely how these are connected remains an outstanding challenge.

We employ two distinct particle dilution methods: reduction of the $f$-electron population by replacing Ce with La, and reduction of conduction electron population by replacing it with non-magnetic Cd or Zn \cite{Pham2006,Yokoyama2014}. Our primary result is that conduction-site hole-doping (cadmium or zinc) acts as a large effective magnetic field in the magnetotransport response of the strange metal, yet has almost no effect on the superconducting transition temperature. By contrast, lanthanum doping strongly suppresses superconductivity, but has a negligible effect on the magnetotransport properties at comparable concentrations. We argue that these findings strongly suggest that the origin of the strange metal behavior is in the degree of Kondo entanglement as a function of temperature.




$\text{CeCoIn}_{5}$ displays a linear-in-T resistivity starting at about 20K accompanying the gradual onset of $f/c$ hybridization with decreasing temperature~\cite{Petrovic2001,Zhou2013,Movshovich2002,Hewson1997}. (Fig. \ref{fig:res} (a,b)) With increasing doping of each type, the zero-field resistivity curve shifts vertically by a constant residual due to static disorder introduced by dopant impurities (Fig. 1 (d))~\cite{Ashcroft1976}; the shape of $\rho(\text{T})$ also changes slightly with doping. Notably, the superconducting transition temperature is most strongly affected by lanthanum doping (Fig. \ref{fig:res} (e)).

\section*{FIGURES}
\begin{figure*}[!htbp]
\centering
\includegraphics[scale=0.75]{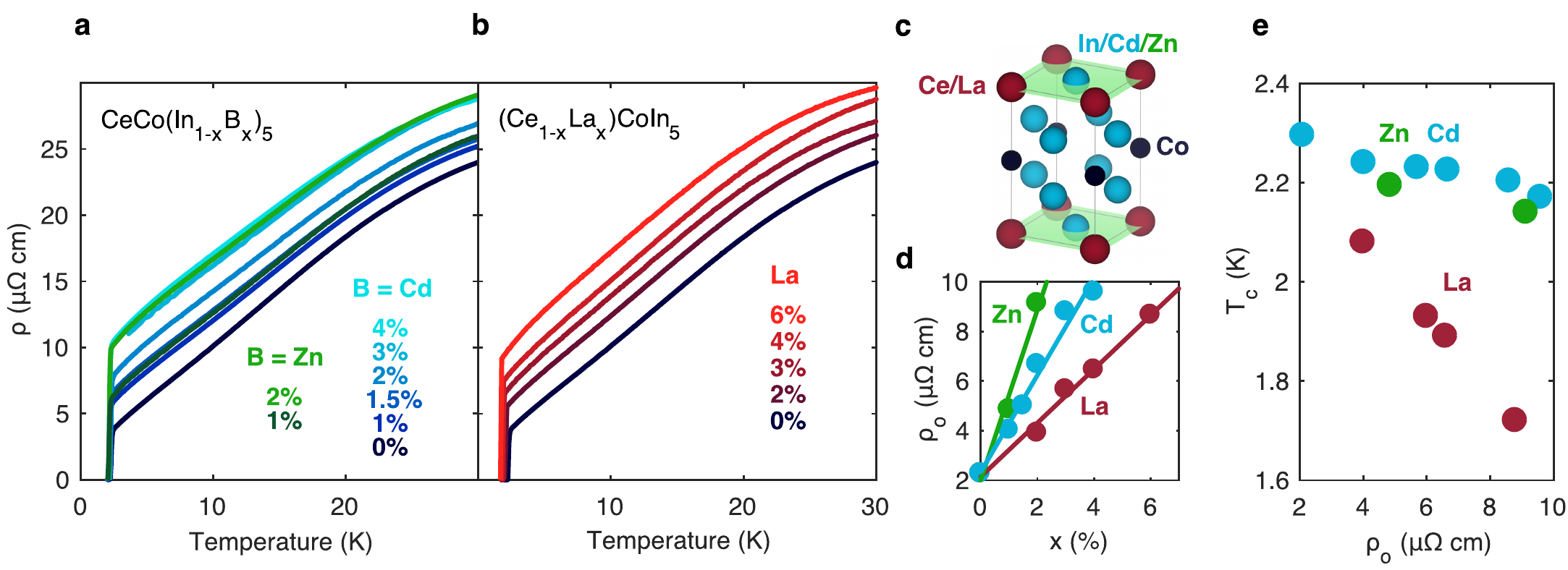}
\caption{{\bf Transport and superconductivity as a function of doping.} (a,b) Resistivity versus temperature at zero applied magnetic field for cadmium/zinc doped samples (blue/green), and lanthanum doped samples (red) respectively. (c) Tetragonal crystal structure of $\text{CeCoIn}_{5}$ consisting of stacked $\text{CeIn}_{3}$ planes, highlighted by light green isosurfaces, separated by layers composed of cobalt and indium. (d) Residual resistivity, extracted using a linear fit to $\rho(\text{T})$ over the 9-14$\text{K}$ temperature range. (e) Superconducting transition temperature versus residual resistivity for each doped sample.}
\label{fig:res}
\end{figure*}

\begin{figure}[!htbp]
\centering
\includegraphics[scale=0.75]{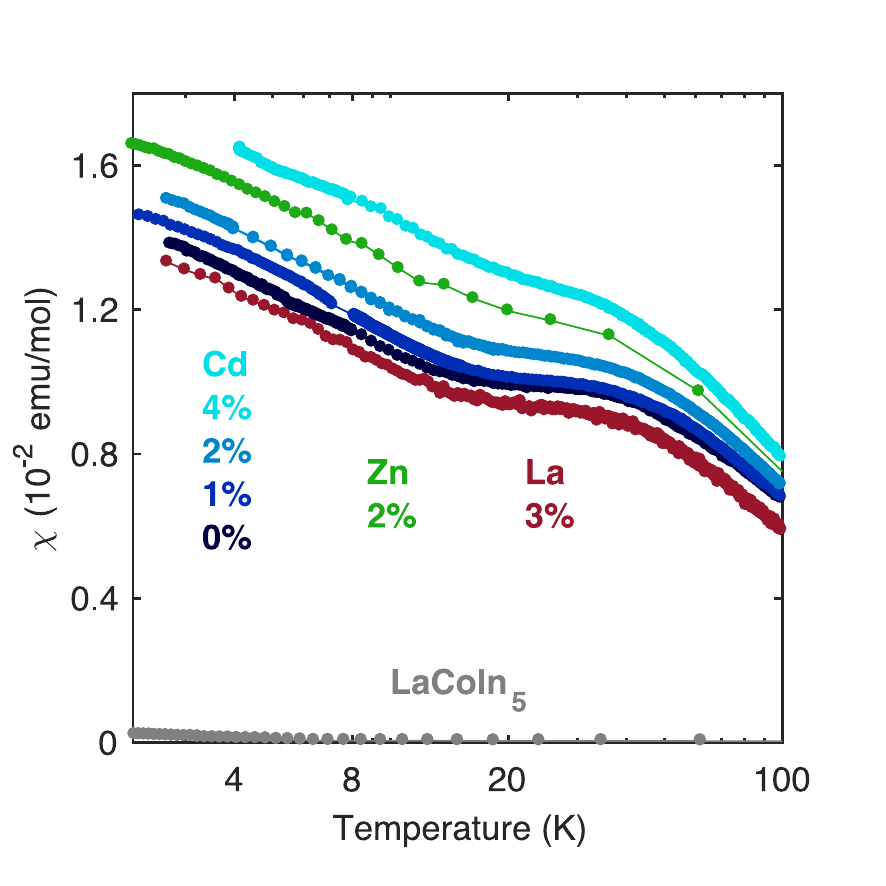}
\caption{{\bf Magnetization of doped samples measured in 50mT} ($\chi = M/H$) along the $c$-axis for $\text{LaCoIn}_{5}$ and various doped samples of $\text{CeCoIn}_{5}$.}
\label{fig:Mag}
\end{figure}


The magnetic susceptibility can be used as a measure of the degree of Kondo compensation. Partially unquenched local 4$f$-moments contribute a magnetic response depending on the degree of Kondo screening~\cite{Meyer2000,Hewson1997}.

In this material, $\chi(T)$ begins to saturate near the Kondo coherence temperature ($T_{coh} \approx 45K$)~\cite{Petrovic2001}, where Kondo screening onsets (Fig. \ref{fig:Mag}). The intrinsic 4$f$ susceptibility contribution in the presence of Kondo compensation can be extracted using $\chi$ below the onset of coherence~\cite{Hewson1997}. From our data, this 4$f$ contribution increases monotonically with conduction-site hole doping (Fig. \ref{fig:Mag}). This indicates that the mean-field hybridization is weakened by conduction-site hole-doping, in agreement with previous literature~\cite{Pepin2007,Meyer2000}. Correspondingly, 4$f$ dilution decreases susceptibility overall. We note that the low-T paramagnetic tail is largely unaffected by dilute doping other than a temperature-independent shift.

\begin{figure*}
\centering
\includegraphics[scale=0.75,trim=0cm 0cm 0cm 0cm]{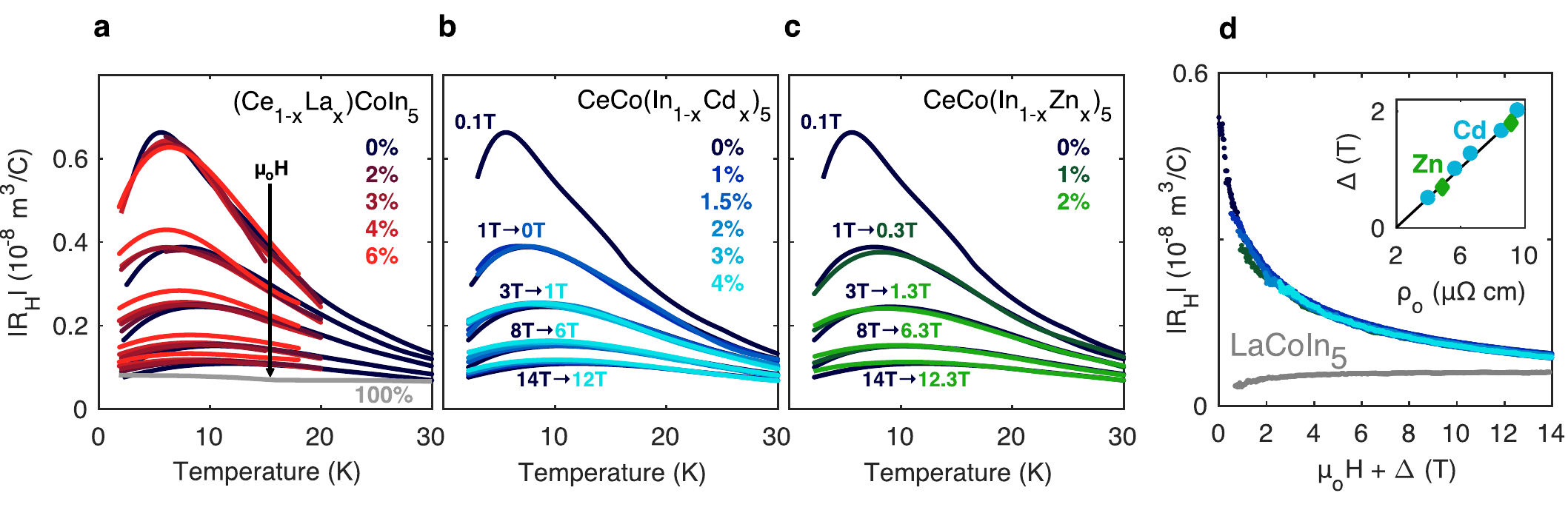}
\caption{{\bf Temperature and field dependence of the Hall coefficient.} $R_{H} (= \rho_{xy}/B$) measured at various fixed fields in (a) La-doped $\text{CeCoIn}_{5}$ at applied fields of 0.1, 1, 3, 8, 14T. For LaCoIn$_5$ (grey line), only the 14T curve is shown. (b,c) $|\text{R}_{\text{H}}|$ for Cd-doped and Zn-doped $\text{CeCoIn}_{5}$. Curves are labeled by the range of applied magnetic fields for different doping concentrations. For example, the $3\text{T} \rightarrow 1\text{T}$ curve in (b) contains data for 0\%, 1\%, 1.5\%, 2\%, 3\%, 4\% cadmium-doped samples measured with applied fields of 3T, 2.5T, 2T, 1.6T, 1.4T, and 1T respectively --- all of these align to the same curve. (d) Field-dependence of the Hall coefficient at a characteristic temperature (10K), containing data for the zinc and cadmium doped samples with the same color scheme as the other panels, as well as $R_{H}(B)$ in $\text{LaCoIn}_{5}$ (grey line). When shifted by the doping-dependent constant $\Delta$, the field-dependence of $R_{H}$ in Cd/Zn-doped samples aligns to the same curve. Inset shows $\Delta(\text{x})$ is a single function of the residual resistivity induced by either cadmium or zinc dopants.}
\label{fig:Hall}
\end{figure*}

We now turn to the effect of particle dilution on transport. In the high-temperature limit (where $f$-electrons do not participate in the Fermi surface~\cite{Chen2017}), $R_{H}$ in CeCoIn$_5$ roughly coincides with that of $\text{LaCoIn}_{5}$, and is independent of magnetic field~\cite{Hundley2004,Nakajima2007}; therefore, before Kondo hybridization occurs on an appreciable scale, $R_{H}$ is approximately consistent with the band structure absent $f$-electrons. Starting near the coherence temperature where $f$-electrons begin to appreciably hybridize with conduction electrons, the zero-field $R_{H}$ remains electron-like, but is strongly enhanced an order of magnitude above that of LaCoIn$_5$ (Fig. \ref{fig:Hall} (a,b,c)). Interestingly, this anomalous temperature-dependence in $R_{H}$ is similar to what is observed in other strange metals~\cite{Abdel-Jawad2007,OFarrell2012,Sun2017,Tyler1997,Taillefer2009}.

In Fig. 3, we plot the Hall effect as a function of temperature for different compositions and magnetic fields. $R_{H}$ in pristine CeCoIn$_5$ is strongly field-dependent, and at high-fields approaches the non-interacting, temperature-independent value of LaCoIn$_5$ (Fig. \ref{fig:Hall} (a,d)). For dilute lanthanum doping, there are only small changes in the temperature or field dependence of $R_{H}$ (Fig. \ref{fig:Hall} (a)). In contrast, cadmium and zinc doping have a strong effect on both the temperature and field dependence of $R_{H}$, systematically decreasing $R_{H}$ and driving the system towards the non-interacting, temperature-independent value (Fig. \ref{fig:Hall} (b,c,d)). Interestingly, the effect of doping directly parallels that of magnetic field over a broad temperature range. The data in doped samples can be mapped onto the field-dependence of the pristine sample with a simple rigid shift in magnetic field, which we label $\Delta(x)$. Symbolically, we write
\begin{equation}
R_{H}(T,B,x) = R_{H}(T,B+\Delta(x),0)
\end{equation}
(Fig. \ref{fig:Hall} (d) inset) For both cadmium and zinc dopants, $\Delta$ is a single function of the residual resistivity, which can be used as a measure of the concentration of dopants~\cite{Ashcroft1976}. Thus, the magnetic field shift, $\Delta$, is approximately linearly proportional to the concentration of incorporated hole-dopants of the conduction sea. (Note that previous alloying studies using zinc and cadmium on $\text{CeCoIn}_{5}$ observe a higher concentration of zinc flux than cadmium flux incorporates into the crystals, equal to about 10\% (for cadmium) and 20\% (for zinc) of the nominal concentration~\cite{Yokoyama2014,Pham2006}). $\Delta$ is unlikely to be the result of a lattice constant effect; one would expect a different function of $\Delta$ versus the residual resistivity for cadmium and zinc, as the residual is only proportional to the concentration of defects~\cite{Ashcroft1976}. This is consistent with a previous cadmium alloying study on this compound, which concludes that the primary effect of this type of doping is likely electronic rather than structural~\cite{Pham2006}.

\begin{figure}[!htbp]
\centering
\includegraphics[scale=0.75]{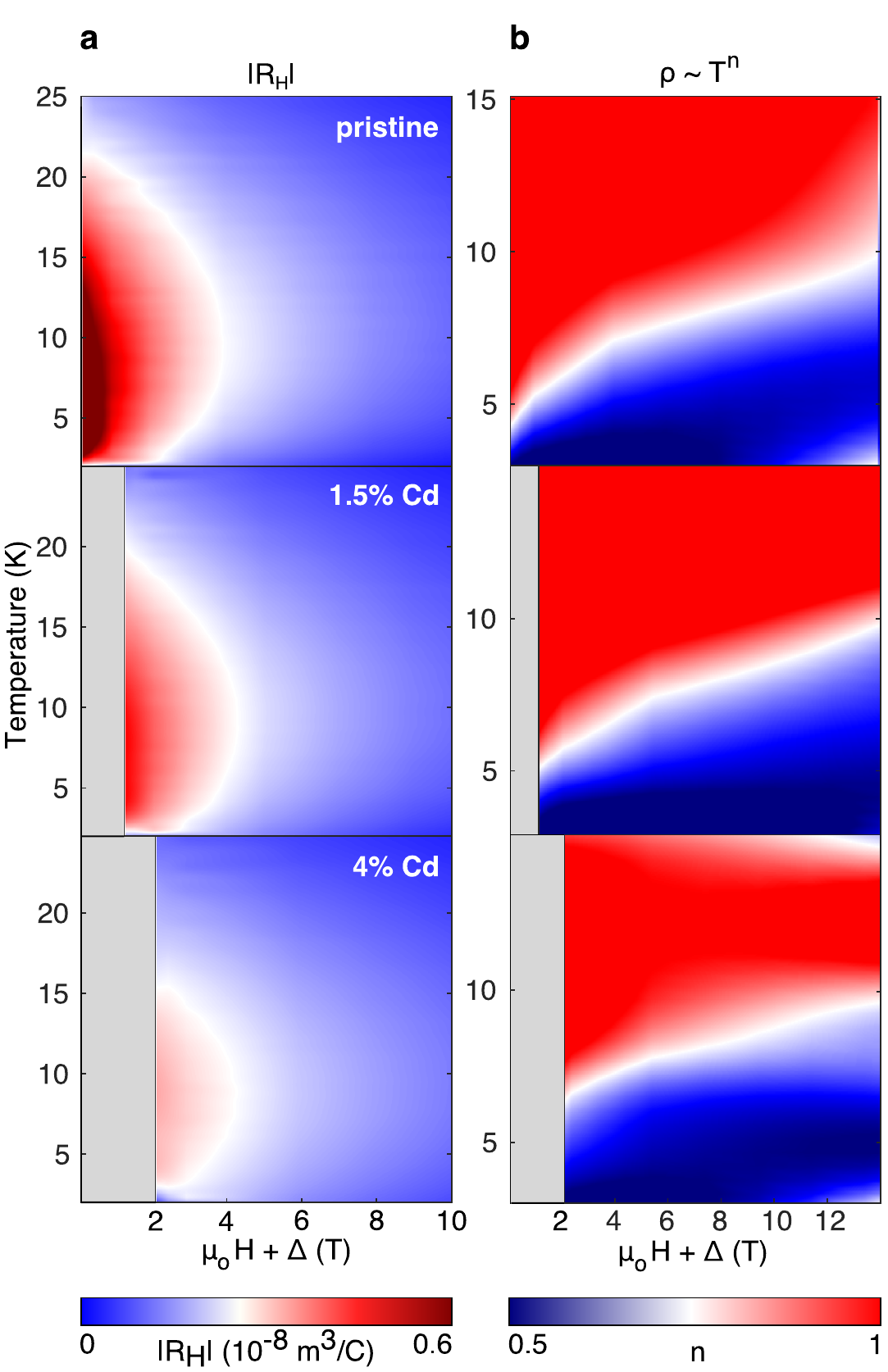}
\caption{{\bf Colomaps of the transport properties as a function of field and temperature.}(a) Color plot of the magnitude of the Hall coefficient as a function of temperature and magnetic field --- shifted horizontally by $\Delta$ in doped samples. (b) Exponent of the temperature-dependence of resistivity, extracted using a least-square power law fit to a 0.5K rolling temperature window to $\rho(T)$ at fixed applied fields. The red ($n=1$) region is the T-linear strange metal regime.}
\label{fig:color}
\end{figure}

From a transport perspective, conduction-site hole-doping rigidly shifts along the field axis of the B-T phase diagram. Fig. \ref{fig:color} summarizes our results, illustrating the application of the $\Delta$-shift rule to the Hall effect (Fig. \ref{fig:color} (a)) and the magnetoresistance (Fig. \ref{fig:color} (b)). The magnitude of $\Delta$ has no correlation to $\text{T}_{c}$ (Fig. \ref{fig:res} (e)); even though conduction-site dopants elicit a strong response in strange metal transport, the pairing mechanism is largely unaffected. In fact, while there is a negligible effect on $\text{T}_{c}$ as a function of conduction-site doping where the $\Delta$-shift rule applies, there is a strong suppression of $\text{T}_{\text{c}}$ as a function of local moment (lanthanum) doping. In the latter, the suppression of $\text{T}_{\text{c}}$ is approximately linear, as expected for large-angle (short wavelength) scattering~\cite{Hamidian2011} in the Abrikosov-Gorkov model~\cite{Abrikosov1975,Zhou2013,VanDyke2014,Petrovic2002}. The change in residual resistivity is comparable for dopants on both local moment and conduction sites, suggesting that the latter has a long-wavelength effect.

Notably, the $\Delta$-shift rule applies to the diagonal magnetoresistance with the same parameter $\Delta$ as for the Hall coefficient (see supplement Fig. 2S (b,c)); algebraically, band mobility enters in a complicated way in the Hall and transverse channels of a multi-band system (having both additive and multiplicative components)~\cite{Pippard2009}. This makes it impossible to capture the observed dependence with a modified $f/c$ mobility and/or carrier number that is independent of $B$ field.

In interpreting the $\Delta$-shift of the Hall effect, we consider the equivalence between $B$-field and conduction electron dilution in this material. LaCoIn$_5$ has the same band structure without $f$-electrons, but has a largely field and temperature independent Hall effect (Fig. \ref{fig:Hall} (a,d)). Note that magnetic field and conduction-site hole-doping equivalently drive the system towards the $R_{H}$ of LaCoIn$_5$. Comparing $R_{H}$ in LaCoIn$_5$ to that of CeCoIn$_5$, it is clear that the Hall coefficient is strongly affected by $f/c$-hybridization below the coherence temperature (Fig. \ref{fig:Hall} (a,d)), though the mechanism of this enhancement is still an open question. Magnetic field gradually eliminates hybridization through the Zeeman interaction~\cite{Hewson1997}, thereby driving $R_{H}$ towards that of LaCoIn$_5$ (Fig. \ref{fig:Hall} (d)). Similarly, decreasing conduction electron density reduces the strength of mean field Kondo coupling (see, e.g.~\cite{Ruhman2014}). This is consistent with the observed increase in magnetic susceptibility (Fig. \ref{fig:Mag})~\cite{Meyer2000,Hewson1997}. Even at dilute doping levels, small changes in the Kondo coupling constant ($J_{K}$) or density of states ($g$) can significantly affect the strength of Kondo hybridization due to the exponential dependence ($\sim \text{exp}(-1/(gJ_{K}))$). In fact, the roughly linear dependence of $\Delta(x)$ likely reflects the linear response of the Kondo coupling to magnetic field~\cite{Hewson1997}, and the first-order linear response to conduction electron dilution. These facts suggest that the temperature-dependence of the Hall coefficient is a direct measure of the strength of mean field Kondo hybridization, which can be weakened with either external $B$-field or decreased conduction electron density.

We now turn to the resistivity, perhaps the most notable of the strange metal phenomena. The $\Delta$-shift in the MR applies to the strange metal temperature range ($< 20 K$), where T-linearity onsets in the pristine sample (Fig. \ref{fig:res} (a)). The equivalent $\Delta$-shift of the resistivity and Hall coefficient provides evidence that the entire resistivity tensor in the strange metal is the consequence of a single underlying mechanism. Namely, strange metal transport in this compound is a direct measure of the strength of Kondo entanglement over a broad temperature range and is irrespective of the process by which the state is disentangled (applied $B$-field or particle dilution). In light of this finding, the most pressing open problem is a precise description of how the many-body hybridized state develops as temperature decreases below the coherence temperature~\cite{Nakatsuji2004,Paglione2007}. This study shows that the transport relaxation of the strange metal is determined directly by the temperature-dependent evolution of this entangled state.


\section*{Acknowledgements}
We would like to thank Snir Gazit, Chandra Varma, and Piers Coleman for helpful discussions. This
work is supported by the Gordon and Betty Moore Foundation’s
EPiQS Initiative through Grant GBMF4374, and by the Office of Naval Research under the Electrical Sensors and Network Research Division, Award No. N00014-15-1-2674.

\section*{Author Contributions}
N.M., B.A., and S.J. grew the crystals used in the experiments. N.M. and I.M.H. performed the transport and magnetization measurements. N.M., E.A. and J.G.A. wrote the manuscript.

\pagebreak
%



\newpage
\section*{Supplement for ``Equivalence of magnetic field and particle dilution in the strange metal state of CeCoIn$_5$"}
\section*{S1 Methods}
Single crystals of $\text{CeCoIn}_{5}$ were grown by an indium self-flux described elsewhere~\cite{Petrovic2001} with a nominal concentration of indium flux replaced by cadmium or zinc, or a nominal concentration of cerium replaced by lanthanum~\cite{Petrovic2001,Pham2006,Yokoyama2014}. Hall bar devices were prepared by mechanically thinning single crystals along the crystallographic $c$-axis to $<$50 $\mu\text{m}$ thickness and attaching gold wires with EpoTek EE129 on gold-sputtered pads. Transverse and Hall resistivity in the $ab$-plane were measured using the standard lockin technique with current excitations between 1-3 mA and magnetic field directed along the crystallographic $c$-axis in a 14T Quantum Design PPMS. Transverse (Hall) resistance was (anti)symmetrized with respect to field polarity. Volume magnetization was measured with a SQUID magnetometer.

Samples are labeled by nominal concentration throughout. Systematic increases in doping concentration in each case were confirmed by a combination of microprobe measurements and a resistive measurement of the superconducting transition temperature in accordance with established literature values~\cite{Pham2006,Kasahara2005,Yokoyama2014}. The concentration of cadmium or zinc dopant inclusions was found to be significantly lower than the nominal concentration in the flux, agreeing with previous Cd or Zn alloying studies on $\text{CeCoIn}_{5}$~\cite{Pham2006,Yokoyama2014}. Doping homogeneity was confirmed by microprobe measurements. Higher doping concentrations (10\%, 20\%) were grown for the cadmium and lanthanum doping series and confirmed by microprobe analysis to lie within the solubility limit.

\section*{S2 Magnetoresistance}
(Fig. 2S (a,b)) $\text{CeCoIn}_{5}$ shows a complicated magnetoresistance (MR), which even in the pristine sample likely reflects the effect of magnetic field on correlations rather than semiclassical Lorentz orbits of quasiparticles~\cite{Paglione2003}. An MR peak can occur in heavy-fermion systems; however, the position of the peak in magnetic field is expected to move to higher fields as the strength of the Kondo condensate increases when temperature is lowered~\cite{Ohkawa1990,Chen1993}. By contrast, when $\text{CeCoIn}_{5}$ displays a T-linear zero-field resistivity, the peak MR field decreases linearly with decreasing temperature~\cite{Paglione2003}. This MR peak as a function of temperature has been attributed in the literature to the delocalization of Kondo particles~\cite{Nakatsuji2004,Paglione2003}.

The MR peak stays at the same field with La-doping, but moves to lower field with increasing Cd/Zn doping (Fig. 2S (a,b)). The same $\Delta$-rule we found for $\text{R}_{\text{H}}$ applies to the transverse MR (Fig. 2S (c,d)). The peak in $\Delta \rho(T)$ at a fixed magnetic field corresponds to a transition from linear-in-T resistivity at moderate temperature to sublinear at low temperature (Fig. 2S (c,d)). Even in the presence of this crossover between scattering regimes, the $\Delta$-shift rule quantitatively holds over about a decade in temperature.

\section*{S3 Skew-scattering considerations}
Skew-scattering mechanisms, which frequently appear in Kondo systems, may seem like natural candidates for understanding our data. While such processes are invariably changing with field or the ratio of $f/c$ electrons, they are likely negligible for the following reasons. Conventional skew-scattering for Kondo systems should be strongest at the coherence temperature~\cite{Coleman1985}, whereas our data show that in this material $R_{H}$ is most strongly enhanced well below the coherence temperature. Moreover, the Hall number does not appear to be related to the magnetization in a simple way (supplement Fig. 1S). For example, La/Ce substitution reduces the number of $f$-electron spins, as evidenced by the decrease in the susceptibility (Fig. \ref{fig:Mag}) and thus should suppress (skew) spin-scattering from $f$-electrons. In contrast, our data shows that $f$-electron dilution has a negligible effect on the Hall coefficient (Fig. \ref{fig:Hall} (a)). This conclusion agrees with previous studies that detail the absence of skew-scattering in this material~\cite{Hewson1997,Nakajima2007}.

\begin{figure}[!htbp]
\centering
   \renewcommand\figurename{FIG. 1S}
\renewcommand{\thefigure}{\hspace{-.333333em}}
\includegraphics[scale=0.8]{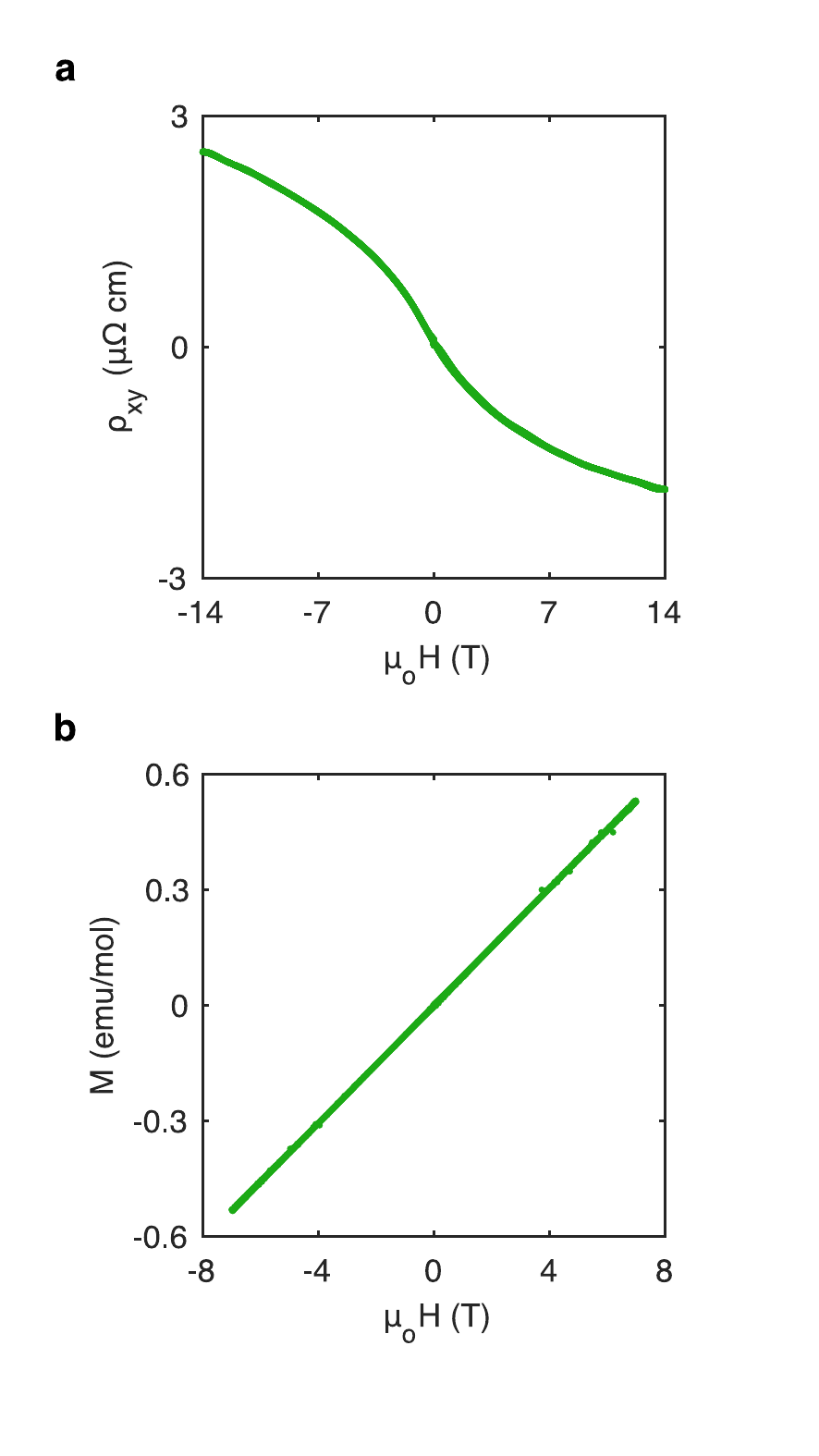}
\caption{(a) Hall effect and (b) Magnetization in a representative 2\% Zn-doped sample at 10 K, including up and down field sweeps. Magnetization is linear in applied field, and neither quantity shows hysteresis.}
\label{fig:hysteresis}
\end{figure}
\vfill\null
\begin{figure*}[!htbp]
\centering
   \renewcommand\figurename{FIG. 2S}
\renewcommand{\thefigure}{\hspace{-.333333em}}
\includegraphics[scale=0.8]{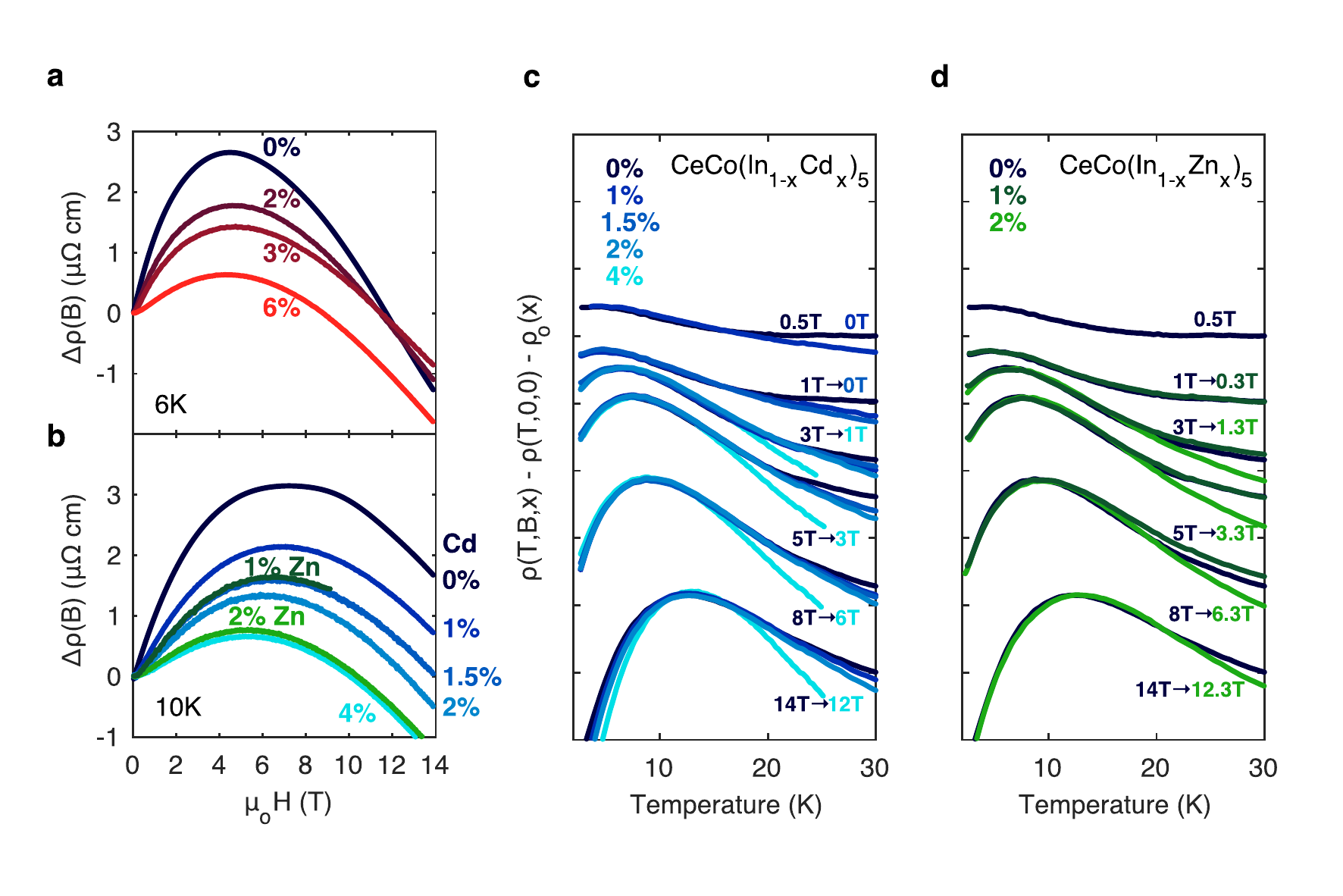}
\caption{(a) Magnetoresistance in lanthanum-doped samples at a representative temperature. (b) Magnetoresistance in cadmium-doped (blue) and zinc-doped (green) samples at 10K. (c,d) Change in $\rho(\text{T})$ as a function of magnetic field and (c) cadmium doping and (d) zinc doping, illustrating deviations from the zero-field, zero-doping T-linear normal state. In addition, the doping-dependent residual, a constant for each sample, was subtracted from the resistivity to eliminate the contribution of elastic impurity scattering. Each set of curves is labeled by the range of applied fields, and offset vertically downwards for clarity. The doped samples in each set merge together under about 20K.}
\label{fig:MR}
\end{figure*}

\end{document}